\documentclass[twocolumn]{autart}    

\usepackage{graphicx}          
\usepackage{empheq}
\usepackage{amssymb}
\usepackage{amsmath}

\newcommand\bR{\mathbb{R}}

\newcommand\cL{\mathcal{L}}

\newcommand\cS{\mathcal{S}}

\newcommand\QED{$~$\hfill{$\square$}}
\newcommand\req[1]{(\ref{eq:#1})}
\newcommand\refig[1]{Fig. \ref{fig:#1}}

\newcommand\reas[1]{Assumption \ref{as:#1}}
\newcommand\reth[1]{Theorem \ref{th:#1}}

\newcommand\relem[1]{Lemma \ref{lem:#1}}

\newcommand\rerem[1]{Remark \ref{rem:#1}}

\newcommand\ressec[1]{\ref{ssec:#1}}
\newcommand\reapp[1]{Appendix \ref{app:#1}}

\renewcommand{\theenumi}{(\roman{enumi})}

\begin{document}

\begin{frontmatter}
\title{Passivity-Based Generalization of Primal-Dual Dynamics for Non-Strictly Convex Cost Functions} 


\author[T2]{Shunya Yamashita}\ead{yamashita.s.ag@hfg.sc.e.titech.ac.jp}
,    
\author[OU]{Takeshi Hatanaka}\ead{hatanaka@eei.eng.osaka-u.ac.jp}               
,    
\author[T2]{Junya Yamauchi}\ead{yamauchi@sc.e.titech.ac.jp}               
,    
\author[T2]{Masayuki Fujita}\ead{fujita@ctrl.titech.ac.jp}               

\address[T2]{Department of Systems and Control Engineering, School of Engineering, Tokyo Institute of Technology
, 
S5-26, 2-12-1 Ookayama Meguro-ku, Tokyo
}  
\address[OU]{Division of Electrical, Electronic and Information Engineering, Graduate School of Engineering, Osaka University
, 
2-1 Yamadaoka, Suita, Osaka
}             

\begin{keyword}                           
Primal-dual dynamics; 
Convex optimization; 
Passivity; 
Distributed optimization;
Invariance principle for Carath\'eodory systems.
\end{keyword}

\begin{abstract}                          
In this paper, we revisit primal-dual dynamics for convex optimization
and present a generalization of the dynamics based on the concept of passivity.
It is then proved that supplying a stable zero to one of the integrators 
in the dynamics allows one to eliminate the assumption of strict convexity on the cost function based on the passivity paradigm together with the invariance principle for Carath\'eodory systems.
We then show that the present algorithm is also a generalization
of existing augmented Lagrangian-based primal-dual dynamics,
and discuss the benefit of the present generalization in terms of noise reduction and convergence speed.
\end{abstract}

\end{frontmatter}

\section{Introduction}

Stimulated by strong needs for solving a large-scale
optimization problem over a spatially distributed network,
primal-dual dynamics \cite{AHU1958}, a continuous-time algorithm to solve convex optimization, has 
attracted attentions again in recent years due to its decomposable nature 
under separability of cost and constraint functions \cite{BV2004}. 
The continuous-time algorithm mitigates the computational efforts, furthermore, it 
does not require network components
to install any optimization solver differently from the other
distributed optimization algorithms \cite{BPet2011}.
Besides, it is pointed out in \cite{WE2011,LQX2014} that
the impact of disturbances and noises added in the optimization process is analyzed from the control engineering point of view,
which is important in the applications to online and/or 
distributed optimization.

The primal-dual dynamics is known to be closely related to
so-called passivity \cite{HCFS2015,K2002}, and it has been revealed that the algorithm
is interpreted as a passivity-preserving interconnection of passive systems \cite{SPS2017,HZet2017,HCDC2017,HCIL2019,YT2012,WA2004}.
The passivity-based perspective brings several advantages.
For example, the design flexibility inherent in passivity-based design 
allows one to stably interconnect other passive components
such as physical dynamics \cite{SPS2017,HZet2017,HCDC2017} and
communication delays with appropriate passivation techniques \cite{HCDC2017,HCIL2019}.
Robustness against the aforementioned disturbances may also be
analyzed based on the celebrated passivity theorem \cite{HCDC2017}.
In addition, the authors in \cite{YT2012,WA2004} point out that 
the design flexibility brought by passivity contributes to
accelerating the convergence speed and/or enhancing robustness.
On the other hand, all of the above papers 
require strict convexity of the cost function, which may limit applications of the solutions.

Relatively few publications have addressed relaxation of the strict convexity assumption
based on so-called augmented Lagrangian.
Richert and Cort\'es \cite{RC2015} present a generalization of the primal-dual dynamics
and prove asymptotic optimality for linear programming problems.
Cherukuri et al. \cite{CDC2017} also present an augmented Lagrangian-based solution to general convex optimization
under strict convexification of the constraint function. 
Zhang et al. \cite{ZWYH2018} present a solution relying on a projection operator
to convex constrained sets although it requires subprocesses to solve optimization to compute the projection.

In this paper, we revisit the paradigm of \cite{YT2012,WA2004}.
We start with hypothesizing that supplying stable zeros to transfer functions,
namely leading the phase,
in the primal-dual dynamics is a key to remove the strict convexity assumption
through a toy linear programming problem.
We then present a passivity-based generalization of the primal-dual dynamics
so that zeros are added to the intended transfer functions.
The above hypothesis is then shown to be valid, namely asymptotic optimality
is proved for general convex cost functions under existence of the zeros, 
based on the passivity paradigm.
It is further demonstrated that the present algorithm is also
a generalization of the existing augmented Lagrangian-based algorithm \cite{RC2015,CDC2017}, 
and the benefit of the generalization is exemplified through simulation.

The major contributions of this paper are summarized as below:
\begin{description}
\item[\textmd{(i)}] 
the passivity-based approaches
\cite{SPS2017,HZet2017,HCDC2017,HCIL2019,YT2012,WA2004}
are extended to general convex optimization with non-strictly 
convex cost function, and 
\item[\textmd{(ii)}] 
a generalization of the augmented Lagrangian-based algorithm \cite{RC2015,CDC2017}
is presented, and the design flexibility inherent in passivity-based design is 
shown to contribute to noise/disturbance reduction and convergence acceleration.
\end{description}
Additional contributions are as follows:
\begin{description}
\item[\textmd{(iii)}] 
the passivity-based generalized primal-dual dynamics 
with general inequality constraints are presented in this paper for the first time, and 
\item[\textmd{(iv)}] 
strict convexification of the constraint function required in \cite{CDC2017} may spoil 
separability, whereas the present approach does not require such reformulation and accordingly 
broadens the class of problems solvable in a distributed fashion. 
\end{description}

\section{Preliminaries}
\label{sec:prelim}

This section is intended to present terminologies, 
associated results and notations used in this paper.

Let us first introduce the notion of passivity
\cite{HCFS2015,K2002}. 
\begin{defn} 
Consider a system $\Sigma$, described by a state model with state $x \in \bR^n$, input $u \in \bR^N$ and output $y \in \bR^N$.  
The system $\Sigma$ is said to be passive if there exists a positive semi-definite function $S : \bR^n \to \bR_{\geq 0} := [0, \infty)$, called storage function, such that 
\begin{align*}
\cL_{\Sigma} S(x) \leq y^\top u 
\end{align*}
holds for all states $x \in \bR^n$ and all inputs $u \in \bR^N$, where
the symbol $\cL_{\Sigma}$ represents Lie derivative along $\Sigma$. 
\end{defn}

We next introduce convex functions defined below. 
\begin{defn} 
A function $f : \bR^n \to \bR$ is said to be convex if the following inequality holds for all $x, y\in \bR^n$. 
\begin{align}
f(x) - f(y) \leq (\nabla f(x) )^\top (x - y) 
\label{eq:def_convex}
\end{align}
\end{defn}
From \req{def_convex}, we immediately have so-called 
monotone condition:
\begin{align}
(\nabla f(x) - \nabla f(y) )^\top (x - y) \geq 0 , ~ \forall x, y \in \bR^n .
\label{eq:ipassive}
\end{align}
If $f$ is strictly convex, the inequality (\ref{eq:ipassive}) strictly holds as long as $x \neq y$ 
\cite{HCIL2019}.

Let us next introduce so-called KKT condition 
for the optimization problem:
\begin{align}
\begin{array}{ccl}
\underset{x \in \bR^n}{\mathrm{minimize}} && f(x)  \\
\mathrm{subject~to} && g(x) \leq 0 , \ A x - b = 0 , 
\end{array}
\label{eq:prob1}
\end{align}
where $x$ is the decision variable, $f : \bR^n \to \bR$ is the 
cost function, $g : \bR^n \to \bR^m$ is the inequality constraint function, $A\in\bR^{r\times n}$ and $b \in \bR^r$ are the constant matrix and vector for equality constraint, respectively. 
Denote the $l$-th element of the function $g$ as $g_l (l = 1, \dots, m) : \bR^n \to \bR$.  
The KKT condition is given as below 
\cite{BV2004}.
\begin{subequations}\begin{align}
& \nabla f(x^*) + \nabla g(x^*) \lambda^* + A^\top \mu^* = 0 ,
\label{eq:kkt1}
\\
& A x^* - b = 0 , 
\label{eq:kkt2}
\\
& \lambda^* \geq 0 , \ \ g(x^*) \leq 0 , \ \  \lambda^* \circ g(x^*) = 0,
\label{eq:kkt3}
\end{align}\label{eq:kkt}\end{subequations}
where the symbol $\circ$ describes the Hadamard product.
The set of the KKT solutions is now defined as
\begin{align}
\chi^* := \left \{ \left . (x^*, \mu^* , \lambda^*) \in \bR^n \times \bR^r \times \bR^m_{\geq 0} \ \right | \req{kkt} \ \rm{holds} \right \} . 
\nonumber
\end{align}

\section{Generalized primal-dual dynamics and passivity}
\label{sec:proposal}

Throughout this paper, we consider the optimization problem
\req{prob1} satisfying the following assumption. 
\begin{assum} \label{as:conv}
The functions $f, g_l \ (l = 1, \dots, m)$ are convex, continuously differentiable, and their gradients $\nabla f, \nabla g_l \ (l = 1, \dots, m)$ are locally Lipschitz. 
The feasible set of \req{prob1} is nonempty, and the function $f$ has a minimum value in the feasible set. 
\end{assum}
It is well-known, under Assumption \ref{as:conv}, that $x^*$ is an optimal solution to \req{prob1} if and only if there exist $(\mu^*, \lambda^*) \in \bR^r \times \bR^m_{\geq 0}$ such that $(x^*, \mu^* , \lambda^*) \in \chi^*$ \cite{BV2004}.
Remark that $x^*$ is not always unique due to the lack of strict convexity of the cost function $f$.

\subsection{Primal-dual gradient dynamics}
\label{ssec:grad}

\begin{figure}
\begin{center}
\includegraphics[width=80mm]{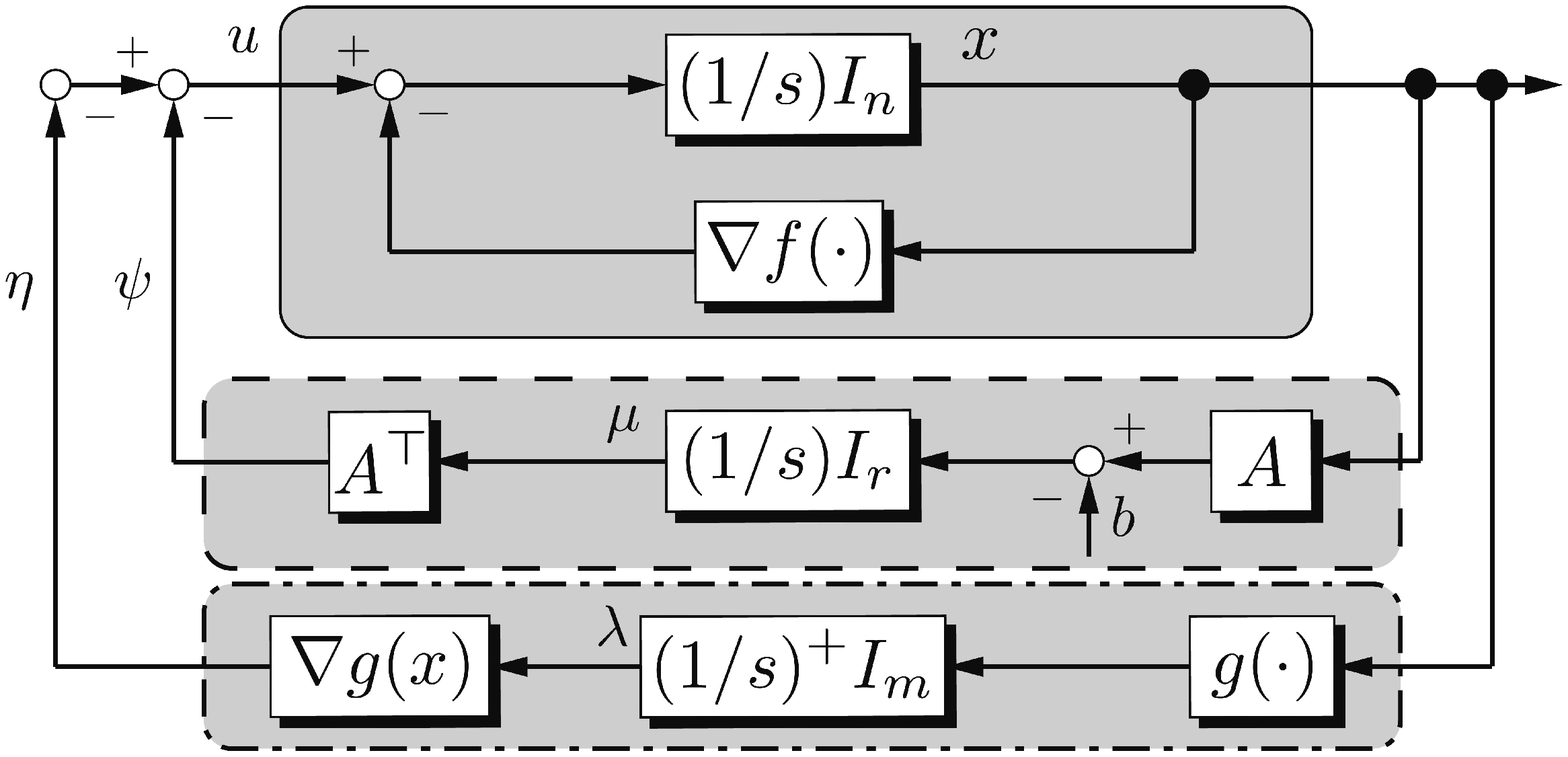}
\caption{Block diagram of the primal-dual gradient dynamics \req{grad}. 
The system enclosed by the solid line is passive from $\widetilde{u}$ to $\widetilde{x}$. 
The system enclosed by the dashed line is passive from $\widetilde{x}$ to $\widetilde{\psi}$. 
The system enclosed by the dashed-dotted line is passive from $\widetilde{x}$ to $\widetilde{\eta}$. 
}
\label{fig:b_grad}
\end{center}
\end{figure}

In this subsection, we first deal with the following primal-dual gradient dynamics \cite{AHU1958,CMC2016} as a solution to \req{prob1}.
\begin{subequations}\begin{align}
\dot{x} &= - \nabla f(x) - \nabla g(x) \lambda - A^\top \mu , 
\label{eq:grad_p}
\\
\dot{\mu} &= Ax - b, 
\label{eq:grad_de}
\\
\dot{\lambda} &= [ g (x) ]_{\lambda}^+ , \ \lambda (0) \geq 0 ,
\label{eq:grad_di}
\end{align}\label{eq:grad}\end{subequations}
where $x \in \bR^n$, $\mu \in \bR^r$ and $\lambda \in \bR^m_{\geq 0}$ are variables corresponding to the primal variable, the dual variable for the equality constraint and the dual variable for the inequality constraint, respectively. 
The operator $[\cdot]^+_{*}$ in \req{grad_di} is defined as
\begin{align}
\left[ \sigma \right]_{\varepsilon}^+ := 
\left \{
\begin{array}{ccl}
0 && \mathrm{if} \ \varepsilon = 0 \ \mathrm{and} \ \sigma < 0 , 
\\
\sigma && \mathrm{otherwise} , 
\end{array}
\right .
\label{eq:ope}
\end{align}
for scalars $\varepsilon, \sigma \in \bR$. 
For vectors $\varepsilon, \sigma \in \bR^N$, $[ \sigma ]^+_{\varepsilon}$ denotes the vector whose $i$-th component is $[\sigma_i]^+_{\varepsilon_i}, \ i = 1, \dots, N $. 
For convenience, the mode satisfying the upper condition in \req{ope} is called mode 1, and the other is mode 2. 
The block diagram of \req{grad} is then illustrated in \refig{b_grad}, where 
\begin{align}
\psi &:= A^\top \mu, \ \eta := \nabla g(x) \lambda, \ 
u := - \eta - \psi.
\nonumber
\end{align} 
and the notation $(\cdot)^+$ means to keep output signal non-negative, defined as below. 
For a transfer function $\frac{1}{\alpha s + \beta}$, the system 
\begin{align}
y(s) = \left ( \frac{1}{\alpha s + \beta} \right )^+ z(s)
\label{eq:op_tfcn}
\end{align}
means that $y(s) = \frac{1}{\alpha s + \beta} z(s)$ under the constraint of $y \geq 0$. 
In other words, if $\alpha \neq 0$, \req{op_tfcn} means 
\begin{align*}
\dot{y} &= \left [ - \frac{\beta}{\alpha} y + \frac{1}{\alpha} z \right ]_{y}^+ ,
\end{align*}
with non-negative initial value $y(0) \geq 0$. 
If $\alpha = 0$ and $\beta > 0$, then \req{op_tfcn} means that 
\begin{align*}
y = \frac{1}{\beta} \max \left \{ 0, z \right \} .
\end{align*}

The primal-dual gradient dynamics \req{grad} is known to satisfy the following facts concerning passivity and convergence 
\cite{HCIL2019,YT2012},
where we take the notations
\begin{align*}
\begin{array}{lcl}
\widetilde{x} := x - x^* , && \widetilde{u} := u - \nabla f(x^*) , 
\\
\widetilde{\psi} := \psi - A^\top \mu^*, && \widetilde{\eta} := \eta - \nabla g(x^*) \lambda^* ,
\end{array}
\end{align*}
for a fixed  $(x^*, \mu^*, \lambda^*)\in \chi^*$.

\begin{fact}
\label{fact:grad}
Suppose that \reas{conv} holds.
Then, the system \req{grad} satisfies the following properties
regardless of the selection of $(x^*, \mu^*, \lambda^*) \in \chi^*$. 
\begin{itemize}
\item The system \req{grad_p} is passive from $\widetilde{u}$ to $\widetilde{x}$,
\item the system \req{grad_de} is passive from $\widetilde{x}$ to $\widetilde{\psi}$,
\item the system \req{grad_di} is passive from $\widetilde{x}$ to $\widetilde{\eta}$, 
\item $(x^*, \mu^*, \lambda^*)$ is a stable equilibrium of the
system \req{grad} in the sense of Lyapunov, and
\item the trajectories of $(x, \mu, \lambda)$ generated by \req{grad} approaches one of the constants included in $\chi^*$ as the time goes to infinity, if the cost function $f$ is strictly convex. 
\end{itemize}
\end{fact}
It is to be noted that only the last item requires strict convexity
of the cost function.
Indeed, the dynamics without this additional assumption
does not ensure asymptotic optimality as exemplified in
the following trivial example.

Let us consider the problem \req{prob1} with
$x\in{\mathbb R}$, $f(x) = 0\ {\forall x}$, $A=1$ and $b=0$ and
without inequality constraints,
which satisfies \reas{conv}.
It is also trivially confirmed that the (unique) optimal solution is $x^* = 0$.
The primal-dual dynamics \req{grad} for the problem 
is then given as
\begin{align}
\dot{x} = - \mu, \ \ 
\dot{\mu} = x.
\label{eq:alg_ex1d}
\end{align}
The dynamics (\ref{eq:alg_ex1d}) is a feedback interconnection
of two single integrators whose open-loop transfer function has 
the phase equal to $-180$deg over the whole frequency domain
and hence the phase margin is $0$deg.
Accordingly, the dynamics does not drive $x$ to $x^* = 0$.

The above toy problem also provides informative knowledge 
in terms of overcoming the drawback of the primal-dual dynamics.
We know that asymptotic stability of $x=0$ for (\ref{eq:alg_ex1d})
is ensured by just adding a compensator leading the phase.
Inspired by the fact, we present a generalization of the 
primal-dual dynamics in the next subsection so that 
the phase lead compensation can be added in this specific example.

\begin{rem}
\label{rem:1}
The primal dual dynamics \req{grad} is known to provide
a distributed algorithm if \req{prob1} is separable
\cite{BV2004}.
In addition, the distributed optimization problem 
\begin{align*}
\underset{x \in \bR^n}{\mathrm{minimize}}& \ \ \sum_{i=1}^N f_i(x)  \\
\mathrm{subject~to}& \ \ g_i(x) \leq 0 , \ A_i x - b_i = 0 , \ {\forall i}=1,\dots, N ,
\nonumber
\end{align*}
with private costs $f_i\ (i=1,\dots, N)$ and private constraints
$g_i(x) \leq 0, \ A_i x - b_i = 0\ (i=1,\dots, N)$
for agents $i=1,\dots, N$
connected by an undirected graph with graph Laplacian $L$
can be equivalently transformed into
\begin{align*}
\underset{x = [x^T_1 \cdots x^T_N]^T \in \bR^{nN}}{\mathrm{minimize}} & \ \sum_{i=1}^N f_i(x_i)  
+ \frac{1}{2} x^\top  (L \otimes I_n) x
\\
\mathrm{subject~to} \quad  & \ g_i(x_i) \leq 0 , \ A_i x_i - b_i = 0 , \ {\forall i}=1,\dots, N , \\
& \ (L\otimes I_n)x = 0 , 
\nonumber
\end{align*}
where the symbol $\otimes$ describes the Kronecker product. 
The primal dual dynamics \req{grad} for the new problem provides
a distributed algorithm based on the PI consensus algorithm
\cite{HCIL2019}. 
These distribution of primal-dual gradient dynamics \req{grad} is due to the diagonal structure of the integrator matrices in \refig{b_grad}. 
\end{rem}

\subsection{Generalized primal-dual dynamics}
\label{ssec:alg}

In this subsection, we generalize the primal-dual dynamics
mainly to ensure asymptotic optimality 
even in the absence of strict convexity assumption.
For notational simplicity, we define the signals $v$, $h$ and $w$ as
\begin{subequations}
\begin{align}
v &= - \nabla f(x) - \nabla g(x) \lambda - A^\top \mu , 
\label{eq:alg_signv}
\\
h &= A x - b, 
\label{eq:alg_signh}
\\
w &= g(x).
\label{eq:alg_signw}
\end{align}\label{eq:alg_sign}\end{subequations}

Let us present the generalized primal-dual dynamics. 
The basic design policy is to allow one to add the phase lead compensators to the open-loop transfer functions, 
while preserving passivity of the subsystems
colored by gray in Fig. \ref{fig:b_grad}, 
formulated as 
\begin{subequations}\begin{align}
x(s) &= M(s) v(s), 
\label{eq:alg_tfcnx}
\\
\mu(s) &= H(s) h(s), 
\label{eq:alg_tfcnm}
\\
\lambda(s) &= G^+(s) w(s),
\label{eq:alg_tfcnl}
\end{align}\label{eq:alg_p}\end{subequations}
where $M(s)$ and $H(s)$ are the transfer function matrices, $G^+(s)$ is the transfer function matrix with the operator in \req{op_tfcn}. 
Although we might be able to take a more general unstructured form,
we restrict the matrices $M(s)$, $H(s)$ and $G^+(s)$
to the diagonal structure as:
\begin{subequations}\begin{align}
M(s) &= \mathrm{diag} \left( M_1(s), \dots , M_n(s) \right) , 
\label{eq:tf_diagM}
\\
H(s) &= \mathrm{diag} \left( H_1(s), \dots , H_r(s) \right) , 
\label{eq:tf_diagH}
\\
G^+(s) &= \mathrm{diag} \left( G^+_1(s), \dots , G^+_m(s) \right) , 
\label{eq:tf_diagG}
\end{align}\label{eq:tf_diag}\end{subequations}
with
\begin{subequations}\begin{align}
M_i(s) &= \frac{c_{i1}}{s} + \sum_{k=2}^{n_i} \frac{c_{ik}}{s + a_{ik}} + d_i , 
\label{eq:tf_Mi}
\\
H_j (s) &= \frac{\bar{c}_{j1}}{s} + \sum_{q=2}^{r_j} \frac{\bar{c}_{jq}}{s + \bar{a}_{jq}} + \bar{d}_j , 
\label{eq:tf_Hj}
\\
G^+_l (s) &= \left( \frac{\hat{c}_{l1}}{s} \right) ^+ + \sum_{p=2}^{m_l} \left ( \frac{\hat{c}_{lp}}{ s + \hat{a}_{lp}}  \right )^+ + \left ( \hat{d}_l \right )^+,
\label{eq:tf_Gl}
\end{align}\label{eq:tf_}\end{subequations}
where
\begin{align*}
&a_{i n_i} > \cdots > a_{i2} > 0,\ c_{ik} > 0\ (k=1,\dots, n_i),\ d_i \geq 0,
\\
& \bar{a}_{j r_j} > \cdots > \bar{a}_{j2} > 0,\ \bar{c}_{jq} > 0\ (q=1,\dots, r_j),\ \bar{d}_j \geq 0, \\
& \hat{a}_{lm_l} > \cdots > \hat{a}_{l2} > 0,\ \hat{c}_{lp} > 0\ (p=1,\dots, m_l),\ \hat{d}_l \geq 0. 
\end{align*}
It is easy to confirm that (\ref{eq:tf_}) allows one to
add the phase lead compensator to the integrators in the primal dual dynamics \req{grad}.
We also immediately see that 
$M(s)$ and $H(s)$ are passive since $M_i(s)$ and $H_j(s)$
are defined by a parallel connection of passive systems,
which is known to preserve passivity.
More precise descriptions on the issue together with
passivity of $G^+(s)$ will be presented in the next subsection. 

\begin{figure}
\begin{center}
\includegraphics[width=80mm]{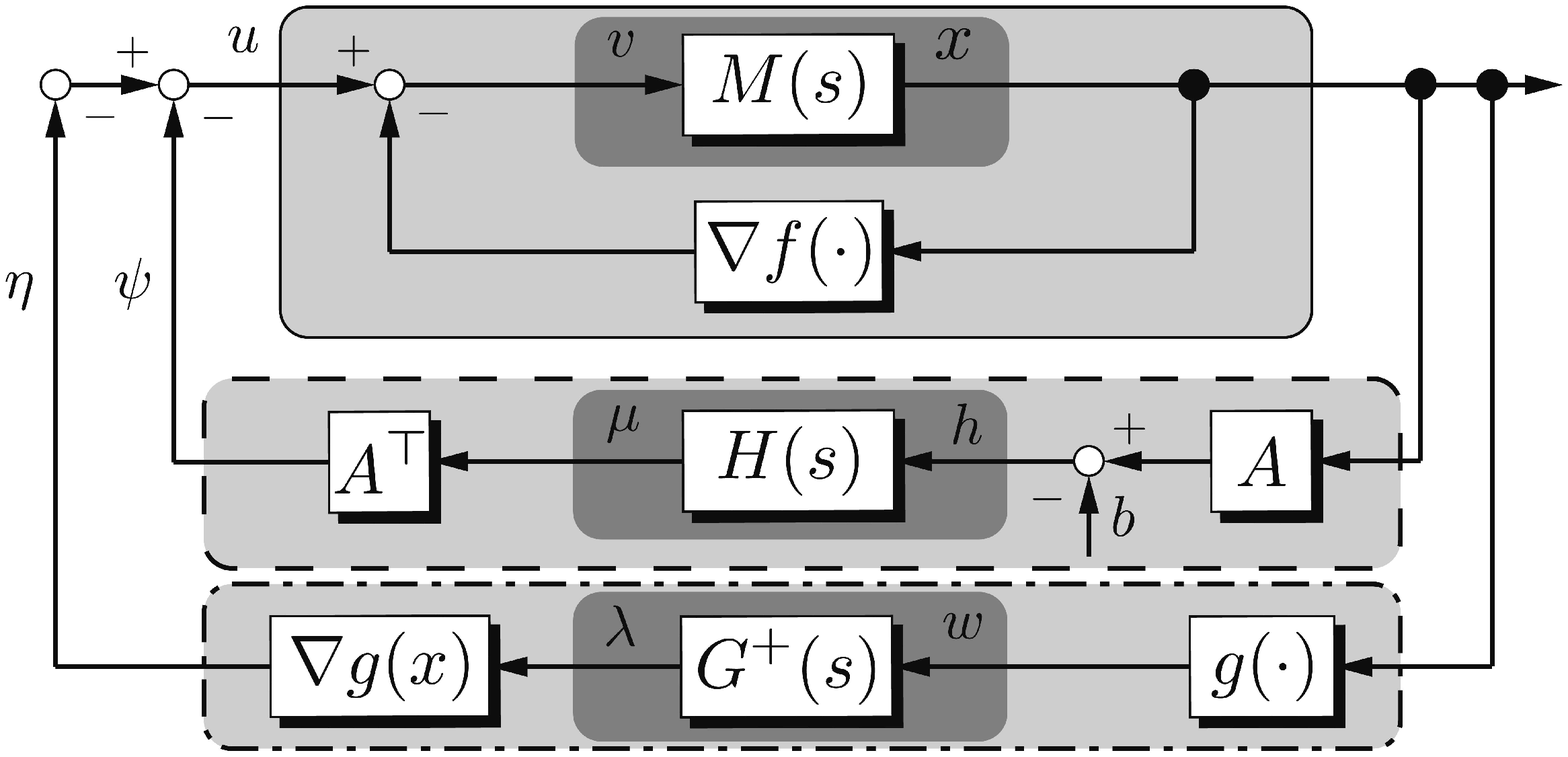}
\caption{Block diagram of the optimization dynamics given by \req{alg_p}--\req{tf_}. 
The dark gray blocks $M(s)$, $H(s)$ and $G^+(s)$ indicate the transfer function matrices given by \req{tf_diag} and \req{tf_}.
The system enclosed by the solid line is passive from $\widetilde{u}$ to $\widetilde{x}$. 
The system enclosed by the dashed line is passive from $\widetilde{x}$ to $\widetilde{\psi}$. 
The system enclosed by the dashed-dotted line is passive from $\widetilde{x}$ to $\widetilde{\eta}$. 
}
\label{fig:b_dc}
\end{center}
\end{figure}

The block diagram of the algorithm \req{alg_p}--\req{tf_} is then illustrated in \refig{b_dc}, where $\psi$, $\eta$ and $u$ are defined in the same way as Subsection \ressec{grad}. 

\begin{rem} 
\label{rem:dist}
At this moment, we focus on the diagonal structure of $M(s)$, $H(s)$ and $G^+(s)$ in (\ref{eq:tf_diag}) for convenience of the subsequent technical discussions, but a more general form will be presented in
the end of the next section. 
However, the diagonal structure is itself of particular importance since it trivially preserves the distributed nature of the primal-dual dynamics pointed out in \rerem{1}.
\end{rem}

\subsection{Passivity analysis}
\label{subsec:pasv}

In this subsection, we confirm that the subsystems
colored by light gray in \refig{b_dc} ensure passivity.

Let us first consider the primal dynamics \req{alg_tfcnx}. 
Now, a state space representation of $M_i(s)$ in \req{tf_Mi} is given as
\begin{subequations}\begin{align}
\dot{\xi}_{i1} &=  c_{i1}v_i ,
\label{eq:statex1}\\
\dot{\xi}_{ik} &= - a_{ik} \xi_{ik} + c_{ik}v_i , \ k = 2, \dots , n_i,
\label{eq:statex2}\\
x_i &= \textbf{1}_{n_i}^\top \xi_i + d_i v_i ,
\label{eq:statex3}
\end{align}\label{eq:statex}\end{subequations}
where $\xi_{ik}$ is the $k$-th element of the state $\xi_i \in \bR^{n_i}$ and
$\textbf{1}_{n_i}$ is the $n_i$ dimensional all-ones vector.
Then, we obtain the following lemma.

\begin{lem} 
\label{lem:pasv_primal}
Suppose that \reas{conv} holds. 
Then, the system given by 
\req{alg_tfcnx}, \req{tf_diagM} and \req{tf_Mi} 
is passive from $\widetilde{u}$ to $\widetilde{x}$ for the following storage function : 
\begin{align*}
S := \sum_{i=1}^{n} S_i , \  S_i := \frac{1}{2c_{i1}} |{\xi}_{i1} - x_i^*|^2 +  \sum_{k=2}^{n_i}  \frac{1}{2c_{ik}}
{\xi}_{ik}^{~2},
\end{align*}
where $x_i^*$ is the $i$-th element of $x^*$.
\end{lem}

\begin{pf}
See \reapp{p_primal}. 
\QED
\end{pf}

We next treat the dynamics \req{alg_tfcnm}. 
A state space representation of $H_j(s)$ in 
\req{tf_Hj} is given as 
\begin{subequations}\begin{align}
\dot{\zeta}_{j1} &= \bar{c}_{j1} h_j ,
\label{eq:statem1}\\
\dot{\zeta}_{jq} &= - \bar{a}_{jq} \zeta_{jq} +\bar{c}_{jq} h_j , \ q = 2, \dots , r_j,
\label{eq:statem2}\\
\mu_j &= \textbf{1}^\top_{r_j} \zeta_j + \bar{d}_j h_j ,
\label{eq:statem3}
\end{align}\label{eq:statem}\end{subequations}
where ${\zeta}_{jq}$ is the $q$-th element of the state $\zeta_j \in \bR^{r_j}$.

\begin{lem} 
\label{lem:pasv_duale}
Suppose that \reas{conv} holds. 
Then, the system given by 
\req{alg_tfcnm}, \req{tf_diagH} and \req{tf_Hj} 
is passive from $\widetilde{x}$ to $\widetilde{\psi}$ for the following storage function : 
\begin{align*}
W := \sum_{j=1}^{r} W_j , \ W_j := \frac{1}{2\bar{c}_{j1}} |\zeta_{j1} - \mu_j^*|^2 + 
\sum_{q=2}^{r_j}  \frac{1}{2\bar{c}_{jq}} {\zeta}_{jq}^{~2},
\end{align*}
where $\mu_j^*$ is the $j$-th element of $\mu^*$.
\end{lem}
\begin{pf}
See \reapp{p_duale}. 
\QED
\end{pf}

Let us finally consider the dynamics \req{alg_tfcnl}. 
The system $G^+_l(s)$ is formulated as follows with
state $\rho_l \in \bR^{m_l}_{\geq 0}$.
\begin{subequations}\begin{align}
\dot{\rho}_{l1} &= [\hat{c}_{l1} w_l]^+_{\rho_{l1}} , 
\label{eq:statel1}
\\
\dot{\rho}_{lp} &= [- \hat{a}_{lp} \rho_{lp} + \hat{c}_{lp}w_l]^+_{\rho_{lp}} , \ p = 2, \dots , m_l ,
\label{eq:statel2}\\
\lambda_l &=\textbf{1}_{m_l}^\top \rho_l + \hat{d}_l \max \{ 0, w_l \} , 
\label{eq:statel3}
\end{align}\label{eq:statel}\end{subequations}
with an initial state $\rho_l (0) \geq 0$, where
${\rho}_{lp}$ is the $p$-th element of $\rho_l$.

\begin{lem} 
\label{lem:pasv_duali}
Suppose that \reas{conv} holds. 
Then, the system given by 
\req{alg_tfcnl}, \req{tf_diagG} and \req{tf_Gl} 
is passive from $\widetilde{x}$ to $\widetilde{\eta}$ for the following storage function : 
\begin{align*}
U := \sum_{l=1}^{m} U_l , \ U_l := \frac{1}{2\hat{c}_{l1}} 
|\rho_{l1} - \lambda^*_l|^2 + 
\sum_{p=2}^{m_l} \frac{1}{2\hat{c}_{lp}} {\rho}_{lp}^{~2},
\end{align*}
where $\lambda^*_l$ is the $l$-th element of $\lambda^*$.
\end{lem}

\begin{pf}
See \reapp{p_duali}. 
\QED
\end{pf}

Lemmas
\ref{lem:pasv_primal}--\ref{lem:pasv_duali} mean that the dynamics \req{alg_p}--\req{tf_} is regarded as a passivity-preserving 
interconnection of passive systems. 
Accordingly, we immediately have the following result
\cite{HCFS2015}. 
\begin{lem} 
\label{lem:7}
Define the function $V := S + W + U$. 
If \reas{conv} is satisfied, 
the Lie derivative along with the system \req{alg_p} of $V$,
denoted by $\cL_A V$, satisfies
$\cL_A V \leq 0$ under \req{alg_p}--\req{tf_} . 
\end{lem}
\begin{pf}
From \req{lem2}, \req{lem4} and \req{lem6}, we obtain 
\begin{align}
\cL_A V &= \cL_P S + \cL_E W + \cL_I U
\nonumber\\
&\leq - \sum_{i=1}^{n} \left ( d_i v_i^2 + \sum_{k=2}^{n_i} \delta_{ik} \xi_{ik}^2 \right ) + \widetilde{x}^\top \widetilde{u} 
\nonumber\\
& \quad \ - \sum_{j=1}^{r}\left ( \bar{d}_j h_j^2 + \sum_{q=2}^{r_i} \bar{\delta}_{jq} \zeta_{jq}^2 \right ) + \widetilde{x}^\top \widetilde{\psi}
\nonumber\\
& \quad \  - \sum_{l=1}^{m}  \left( \hat{d}_l \max \{ 0, w_l \}^2  + \sum_{p=2}^{m_l} \hat{\delta}_{lp} \rho_{lp}^2  \right) + \widetilde{x}^\top \widetilde{\eta}
\nonumber\\
&\leq \widetilde{x}^\top \left( \widetilde{u} + \widetilde{\psi} +  \widetilde{\eta} \right) = 0 ,
\label{eq:lem7}
\end{align}
since $\widetilde{u} + \widetilde{\psi} +  \widetilde{\eta} = 0$ is ensured by \req{kkt1}. 
\QED
\end{pf}


\section{Convergence analysis}
\label{sec:conv}

In this section, we analyze convergence of the trajectories of $(x, \mu,  \lambda)$ generated by \req{alg_p}--\req{tf_} to the set $\chi^*$. 
In the sequel, we use the notations $\bar{n} := \sum_{i=1}^{n} n_i$, $\bar{r} := \sum_{j=1}^{r} r_j$ and $\bar{m} := \sum_{l=1}^{m} m_l$. 
Also, we define the notations of the state variables as 
\begin{align*}
\xi &:= \left[\xi_1^\top \  \dots \  \xi_n^\top \right]^\top \in \bR^{\bar{n}} ,
\\
\zeta &:= \left[\zeta_1^\top \  \dots \  \zeta_r^\top \right]^\top \in \bR^{\bar{r}} ,
\\
\rho &:= \left[\rho_1^\top \  \dots \  \rho_m^\top \right]^\top \in \bR^{\bar{m}}_{\geq 0} .
\end{align*}

The proof relies on the invariance principle 
for Carath\'eodory systems \cite{CMC2016,BC2006}.
It is thus first proved that the system \req{alg_p}--\req{tf_} satisfies the assumptions required by the principle. 
\begin{lem} 
\label{prop:IPP}
Under \reas{conv}, the system \req{alg_p}--\req{tf_} with state $(\xi, \zeta, \rho)$
satisfies the following properties.
\renewcommand{\labelenumi}{(\roman{enumi})}
\begin{enumerate}
\item There exists a compact and invariant subset $\cS \subset \bR^{\bar{n}} \times \bR^{\bar{r}} \times \bR^{\bar{m}}_{\geq 0}$. 
\label{IPP1}
\item For each point $(\xi_0, \zeta_0, \rho_0) \in \cS$, there exists a unique solution of \req{alg_p}--\req{tf_} starting at $(\xi_0, \zeta_0, \rho_0)$. 
\label{IPP2}
\item The omega-limit set of the unique solution of \req{alg_p}--\req{tf_} is invariant. 
\label{IPP3}
\item The Lie derivative of the continuous and differentiable function $V$ along \req{alg_p} satisfies $\cL_A V(\xi, \zeta, \rho) \leq 0$ for all $(\xi, \zeta, \rho) \in \cS$. 
\label{IPP4}
\end{enumerate}
\end{lem}

\begin{pf}
The item (iv) was already proved in Lemma \ref{lem:7}.
The item (i) also immediately holds since
the function $V$ is radially unbounded.
Since \req{alg_p}--\req{tf_} is a projected dynamics,
the item \ref{IPP2} can be proved by following the same procedure as 
Lemma 4.3 in \cite{CMC2016}. 
Additionally, the assumption \ref{IPP3} can be also proved in the same way as 
Lemma 4.4 in \cite{CMC2016} and Lemma 4.1 in \cite{K2002}. 
\QED
\end{pf}

We are now ready to show the main result of this paper.

\begin{thm} 
\label{th:1}
Consider the system \req{alg_p}--\req{tf_}. 
Assume that the transfer function $M_i(s)$ has one or more stable zeros, for all $i$. 
If \reas{conv} holds, then $(x, \mu, \lambda)$ approaches one of the constants included in $\chi^*$ as the time goes to infinity. 
\end{thm}

\begin{pf}
From Lemma \ref{prop:IPP},
the invariance principle for Carath\'eodory systems \cite{CMC2016,BC2006} 
is applied to the system  \req{alg_p}--\req{tf_}, and hence 
any solution of \req{alg_p}--\req{tf_}  starting at $\cS$ converges to the largest invariant set in $\mathrm{cl}(\{ (\xi, \zeta, \rho) \in \cS | \cL_A V(\xi, \zeta, \rho) = 0 \})$. 
From \req{lem7}, $\cL_A V \equiv 0$ implies that 
\begin{align}
& d_i v_i^2 + \sum_{k=2}^{n_i} \delta_{ik} \xi_{ik}^2  \equiv 0 \ \forall i=1, \dots , n , 
\label{eq:equ1}
\\
& \bar{d}_j h_j^2 + \sum_{q=2}^{r_i} \bar{\delta}_{jq} \zeta_{jq}^2 \equiv 0 \ \forall j=1, \dots , r , 
\label{eq:equ2}
\\
& \hat{d}_l \max \{ 0, w_l \}^2 + \sum_{p=2}^{m_l} \hat{\delta}_{lp} \rho_{lp}^2 \equiv 0 \ \forall l=1, \dots , m. 
\label{eq:equ3}
\end{align}

In the sequel, we consider the system trajectories identically satisfying
(\ref{eq:equ1})--(\ref{eq:equ3}).
First, we focus on \req{equ1} and the dynamics (\ref{eq:statex}). 
Since $M_i(s)$ has one or more zeros, $d_i > 0$ or $n_i \geq 2$ must be satisfied. 
If $d_i > 0$, we have $v_i \equiv 0$ and, otherwise, we have
$\xi_{ik} \equiv 0$ and $\dot{\xi}_{ik} \equiv 0$ for all $k=2,\dots , n_i$. 
In the latter case, substituting $\xi_{ik} \equiv 0$ and $\dot{\xi}_{ik} \equiv 0$ into \req{statex2} yields $v_i \equiv 0$.
We thus conclude that $v_i \equiv 0$ holds for all $i$, and hence 
the state trajectories must identically satisfy
\begin{align}
\nabla f(x) + \nabla g(x) \lambda + A^\top \mu = 0.
\label{eq:cov_kkt1}
\end{align}
We also see from \req{statex1} and
\req{statex3} that $\xi_{i1}$ and $x_i$ must be constant for all $i$.

Next, we focus on \req{equ2} and the dynamics (\ref{eq:statem}). 
Since $x$ is constant as shown above, $h = Ax - b$ must be also constant. 
Now, if $h_j \neq 0$ for some $j$, ${\zeta}_{j1}$ must diverge from \req{statem1},
which contradicts boundedness of $\zeta_{j1}$.
We thus conclude that the trajectories meet $h_j \equiv 0$. 
In other words, the following equation identically holds. 
\begin{align}
A x - b = 0
\label{eq:cov_kkt2}
\end{align}
Accordingly, $\zeta_{j1}$ is constant.  
We also have $\zeta_{jq} \equiv 0 \ \forall q = 2, \dots, r_j$ from \req{equ2}.
In summary, we conclude from \req{statem3} that 
the trajectories satisfying
(\ref{eq:equ1})--(\ref{eq:equ3}) meet
$\mu_j \equiv \bar{c}_{j1} \zeta_{j1}$ 
and it is identically constant.

Let us next consider about \req{equ3} and the dynamics \req{statel}. 
Since $x$ is constant as shown above, $w = g(x)$ is also constant. 
Now, let us focus on \req{statel1}. 
If mode 1 is active, then $\dot{\rho}_{l1} = 0$ holds. 
Otherwise, $\dot{\rho}_{l1} = w_l$ holds, i.e., $\dot{\rho}_{l1}$ is constant. 
Then, $\dot{\rho}_{l1} = 0$ holds since $\dot{\rho}_{l1} \neq 0$ contradicts the boundedness of $\rho_{l1}$. 
Thus, we have $\dot{\rho}_{l1} \equiv 0$ for all $l = 1, \dots, m$. 
This means that 
\begin{align}
\rho_{l1} \geq 0 , \ w_l \leq 0 , \ \rho_{l1} w_l = 0 , \ \forall l = 1, \dots, m 
\label{eq:cov_kktr}
\end{align}
is identically satisfied. 
From \req{equ3}, we obtain 
\begin{align*}
\hat{d}_l \max \{ 0, w_l \} \equiv 0 , 
\ \ \ 
\rho_{lp} \equiv 0 , \ \forall p = 2, \dots, m_l .
\end{align*}
From \req{statel3}, we also see that $\lambda_l \equiv \hat{c}_{l1} \rho_{l1}$
and it is identically constant. 
Moreover, (\ref{eq:cov_kktr}) means that the trajectories identically satisfy
\begin{align}
\lambda \geq 0, \ g(x) \leq 0 , \ \lambda \circ g(x) = 0 .
\label{eq:cov_kkt3}
\end{align}

In summary, we conclude that the state trajectories identically satisfying
(\ref{eq:equ1})--(\ref{eq:equ3}) provide a constant
$(x, \mu, \lambda)$ satisfying \req{cov_kkt1}, \req{cov_kkt2} and \req{cov_kkt3}. 
In other words, $(x, \mu, \lambda) \in \chi^*$ holds in the positively invariant set in $\cL_A V \equiv 0$. 
This completes the proof.
\QED
\end{pf}
Remark that the assumption on the zeros of $M_i(s)$ validates the hypothesis extracted from the toy problem in the end of the previous section that leading the phase is the key to relax the strict convexity assumption.

Similar generalizations of the primal-dual dynamics are presented \cite{YT2012,WA2004}.
In these publications, $M_i(s)$ is assumed to be proper, positive real and having a pole at the origin, which is almost compatible with ours, and asymptotic optimality is proved under strict convexity of the cost function. 
The primary contribution of this paper relative to \cite{YT2012,WA2004} is to show that the strict convexity assumption can be relaxed to convexity under the additional condition on the zeros of $M_i(s)$. 
Besides, we can add two additional contributions as below.

First, our algorithm can treat a general convex constraint function $g$, while \cite{YT2012} and \cite{WA2004} deal with the problems without inequality constraints and with linear inequality constraints, respectively. 
Namely, we immediately have a fully generalized result of \cite{YT2012,WA2004} as follows.
\begin{cor} 
\label{cor:st_conv}
Consider the system \req{alg_p}--\req{tf_}. 
Suppose that \reas{conv} holds. 
If the cost function $f$ is strictly convex, $(x, \mu, \lambda)$ approaches one of the constants included in $\chi^*$ as the time goes to infinity. 
\end{cor}
\begin{pf}
Noticing \req{ineq_l2_2} and \req{lem7}, $\cL_A V \equiv 0$ implies  
\begin{align}
(x - x^*)^\top (\nabla f (x) - \nabla f(x^*)) \equiv 0 .
\label{eq:equ_st1}
\end{align}
Due to strict convexity of $f$, \req{equ_st1} is equivalent to $x \equiv x^*$. 
Also, $x$ is constant because $x^*$ is the unique solution to \req{prob1}. 
From these results, we can prove that $\mu$ and $\lambda$ are constant in the same way as \reth{1}. 
In addition, we obtain that $x$ satisfies \req{cov_kkt2} and \req{cov_kkt3}. 
Since $x$, $\mu$ and $\lambda$ are constant, $v = -\nabla f(x) - \nabla g(x) \lambda - A^\top \mu$ is also constant. 
Then, we have $v \equiv 0$ since $\dot{\xi}_{i1} = v_i \neq 0$ contradicts the boundness of $\xi$. 
Thus, \req{cov_kkt1} holds. 
As a result, $(x, \mu, \lambda)$ is the constant satisfying KKT conditions when $\cL_A V \equiv 0$. 
This completes the proof. 
\QED
\end{pf}

In addition, \cite{YT2012,WA2004} take the diagonal transfer function matrices $M(s)$ and $H(s)$ as in (\ref{eq:tf_diag}) and we can also generalize the structure based on the passivity paradigm. 
To this end, we first replace $M(s)$ and $H(s)$ in (\ref{eq:tf_diag}) by 
\begin{subequations}\begin{align}
M(s) &= \mathrm{diag} \left( M_1(s), \dots , M_n(s) \right) + M'(s), 
\label{eq:tf_diagM2} 
\\
H(s) &= \mathrm{diag} \left( H_1(s), \dots , H_r(s) \right) + H'(s), 
\label{eq:tf_diagH2}
\end{align}\label{eq:tf_diag2}\end{subequations}
where $M'(s)$ and $H'(s)$ are possibly non-diagonal transfer function matrices assumed to be strictly positive real. 
The additional design flexibility associated with $M'(s)$ and $H'(s)$ may contribute to improvement of the performance.
It is now not difficult to confirm that adding $M'(s)$ and $H'(s)$ does not affect all the signals at the stationary state.
From passivity preservation w.r.t. parallel interconnections, both of $M(s)$ and $H(s)$ are preserved to be passive. 
Accordingly, we can immediately prove the following corollary. 
\begin{cor} 
\label{cor:ff_prf}
Consider the system \req{alg_p}, \req{tf_diag2}, \req{tf_diagG} and \req{tf_}. 
Assume that the transfer function $M_i(s)$ has one or more stable zeros, for all $i$. 
If \reas{conv} holds, then $(x, \mu, \lambda)$ approaches one of the constants included in $\chi^*$ as the time goes to infinity. 
\end{cor}
\begin{pf}
Redefine the energy function $V$ by adding the storage functions of $M'(s)$ and $H'(s)$.
It is then immediate to see from passivity preservation of $M(s)$ and $H(s)$ that
the inequality (21) holds even after adding $M'(s)$ and $H'(s)$.
The subsequent discussions are the same as \reth{1}.
\QED
\end{pf}

\section{Relation to augmented Lagrangian method}
\label{sec:augL}

In this section, we explore relations between the present algorithm \req{alg_p} and augmented Lagrangian-based primal-dual dynamics
in \cite{RC2015,CDC2017}.

\begin{figure}
\begin{center}
\includegraphics[width=80mm]{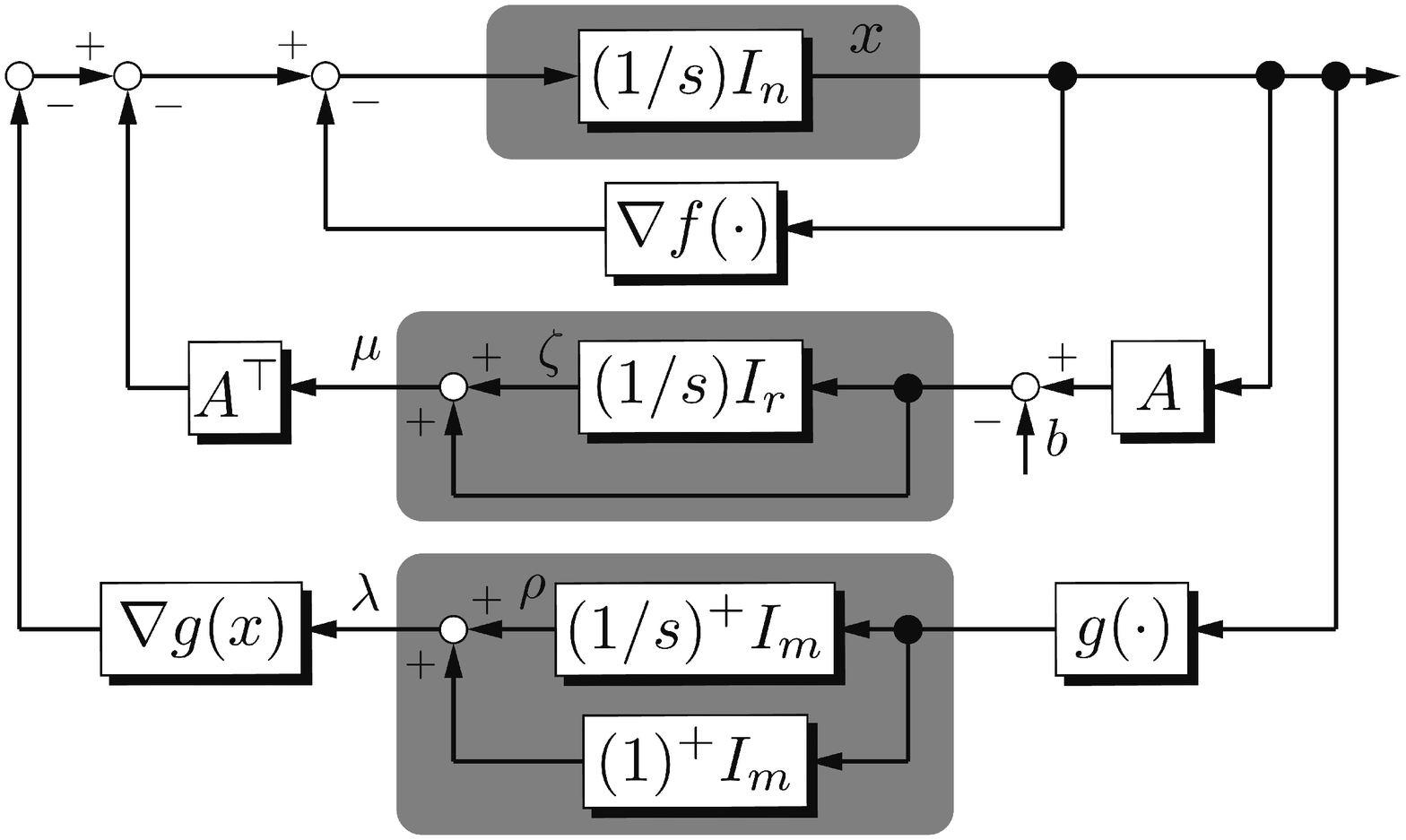}
\caption{Block diagram of the dynamics \req{alg_aL1}. 
This dynamics is a special case of the system in \refig{b_dc} with $M_i (s) = \frac{1}{s}$, $H_j (s) = \frac{s + 1}{s}$ and $G^+_l(s) = \left( \frac{1}{s} \right)^+ + (1)^+ $. 
}
\label{fig:block_diag1}
\end{center}
\end{figure}

Let us first focus on \cite{CDC2017}, where 
the authors present the following dynamics to solve (\ref{eq:prob1}).
\begin{subequations}\begin{align}
\dot{x} &=  - \left( \nabla f(x) + \nabla g(x) \lambda + A^\top \mu  \right) , 
\\
\mu &= \zeta + (Ax - b) , 
\\
\dot{\zeta} &= Ax - b ,
\\
\lambda &= \rho + \max \{ 0, g(x) \} , 
\\
\dot{\rho} &=  [ g (x) ]^+_{\rho} , \ \rho (0) \geq 0 .
\end{align}\label{eq:alg_aL1}\end{subequations}
It is not difficult to confirm that (\ref{eq:alg_aL1}) is illustrated in
the block in Fig. \ref{fig:block_diag1}.
Comparing Fig. \ref{fig:block_diag1} with Fig. \ref{fig:b_dc},
we immediately see that (\ref{eq:alg_aL1}) is a special case of
\req{alg_p}.
Specifically, if we take $M_i(s) = \frac{1}{s}$, $H_j(s) = \frac{s + 1}{s}$ and $G_l^+(s) = \left( \frac{1}{s} \right)^+ + \left( 1 \right)^+$,
\req{alg_p} coincides with \req{alg_aL1}. 
We thus conclude that the present dynamics is a generalization of
\req{alg_aL1}.

\begin{figure}
\begin{center}
\includegraphics[width=80mm]{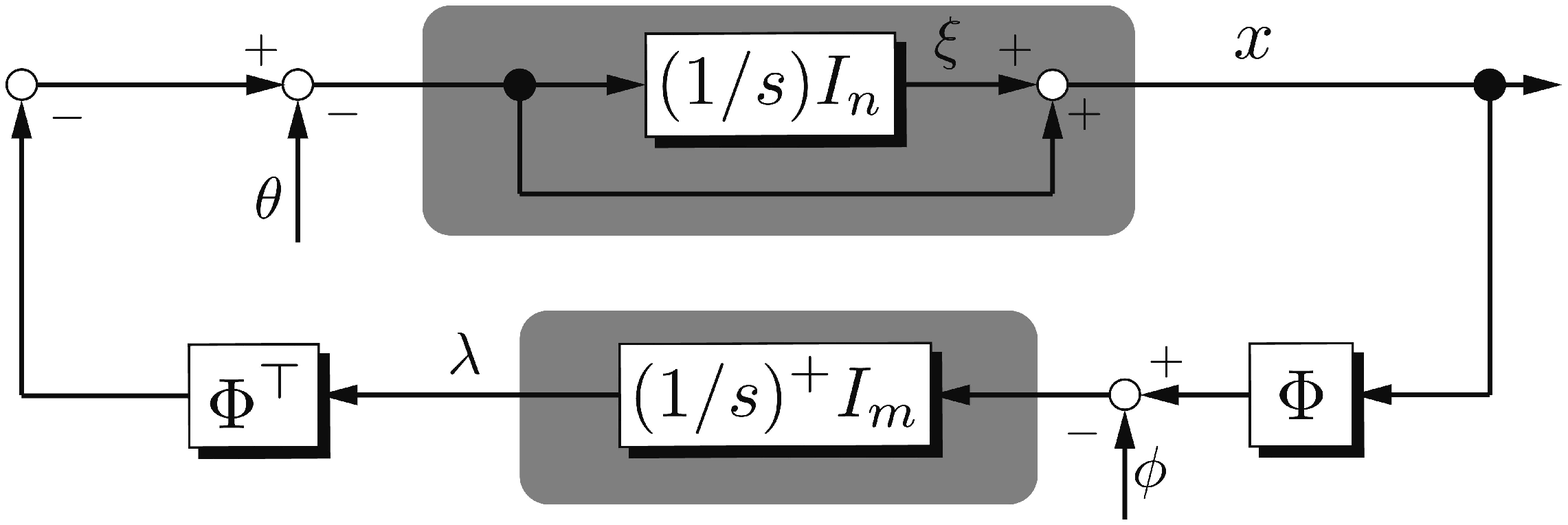}
\caption{Block diagram of the dynamics \req{alg_aLlp2}. 
This dynamics is a special case of the system in \refig{b_dc} with $M_i(s) = \frac{s + 1}{s}$ and $G^+_l (s) = \left( \frac{1}{s} \right)^+ $. 
}
\label{fig:block_diag2}
\end{center}
\end{figure}

We next investigate the relation to
\cite{RC2015}, where 
the authors address the following linear programming problem.
\begin{align}
\begin{array}{ccl}
\underset{x \in \bR^n}{\mathrm{minimize}} && \theta^\top x , \\
\mathrm{subject~to} && \Phi x - \phi \leq 0 ,
\end{array}
\label{eq:lp1}
\end{align}
where $\Phi \in \bR^{m \times n}$, $\phi \in \bR^m$ and $\theta \in \bR^n$. 
Then, the following dynamics to solve (\ref{eq:lp1})  
is presented in \cite{RC2015}.
\begin{subequations}\begin{align}
x &= \xi - \theta - \Phi^\top \lambda, 
\\
\dot{\xi} &= - \theta - \Phi^\top \lambda, 
\\
\dot{\lambda} &= \left[ \Phi x - \phi  \right]^+_{\lambda} , \ \lambda (0) \geq 0 ,
\end{align}\label{eq:alg_aLlp2}\end{subequations}
which is illustrated in Fig. \ref{fig:block_diag2}.
We immediately see from the figure that \req{alg_aLlp2} is equivalent to
\req{alg_p} with
$M_i(s) = \frac{s + 1}{s}$ and $G_l^+(s) = \left( \frac{1}{s} \right)^+$.
It is thus concluded that \req{alg_aLlp2} is also a special example of \req{alg_p}.

In the reminder of this section, we clarify benefits of the generalized algorithm \req{alg_p} over \cite{RC2015,CDC2017}.
While the advantage over \cite{RC2015} is obvious since our approach is not restricted to the linear programming,
the result of \cite{CDC2017} looks compatible with ours.
However, \cite{CDC2017} requires an additional assumption,
namely strict convexity of the constraint function $g(x)$, 
in order to prove asymptotic convergence to the optimal solution. 
Now, even if the original constraint function is separable,
the strictly convexified function may lose the separable structure,
which can be an obstacle for distributed optimization.
Meanwhile, the present approach does not require such operations.

As stated in \cite{RC2015,SPMD2018}, optimization algorithms may suffer from a variety of noises. For example, in online optimization, the cost and constraint functions may be defined by the real-time data including noises. 
In addition, distributed implementation of the algorithm may suffer from the noises at the communication channels.
Regarding the noise reduction, it is to be noted that only one of the transfer functions $M_i(s)$, $H_j(s)$ and $G_l^+(s)$ in \cite{RC2015,CDC2017}
are strictly proper, which means that the gain decay of the open-loop systems over the high frequency domain is 20dB/dec.
On the other hand, our approach allows one to choose
strictly proper $M_i(s)$, $H_j(s)$ and $G_l^+(s)$, which would
achieve a better roll-off and hence better noise reduction.
Besides, the generalization presented in this paper allows one to shape the open loop systems more flexibly, which would contribute to a better disturbance rejection and/or acceleration of convergence speed.

We exemplify the above hypothesis through simulation.
Let us consider the linear programming problem \req{lp1} with 
\begin{align*}
\Phi = \left[ \begin{array}{cc}
-1 & 0\\
0 & -1\\
4 & 3\\
1 & 2
\end{array}
\right] ,\ \phi = \left[ \begin{array}{c}
0\\
0\\
10\\
5
\end{array}
\right] , \ \theta = \left[ \begin{array}{c}
-2\\
-3
\end{array}
\right]
\end{align*}
whose optimal solution is $x^* = [1 \ 2]^\top$. 
We prepare the following three dynamics to solve the problem.  
\begin{itemize}
\item Case 1 (\cite{RC2015}) : 
\\
$M_i(s) = \frac{s+1}{s}  ,\ \forall i, \ \ G_l^+ (s) = \left( \frac{1}{s} \right)^+, \ \forall l$
\item Case 2 : 
\\
$M_i(s) = \frac{1}{s} + \frac{19}{s + 25} , \ \forall i , \ \ G_l^+ (s) = \left( \frac{1}{s} \right)^+, \ \forall l$
\item Case 3 : \\
$M_i(s) = \frac{1}{s} + \frac{19}{s + 25} , \ \forall i , \ \ G_l^+ (s) = \left( \frac{1}{s} \right)^+ + \left( \frac{4}{s + 0.05} \right)^+, \ \forall l$
\end{itemize}
Note that all of them satisfy the assumptions in \reth{1}, and 
case 1 coincides with the algorithm in \cite{RC2015}.

\begin{figure*}[!t]
\begin{center}
\begin{tabular}{cc}
  \begin{minipage}{.30\hsize}
    \begin{center}
\includegraphics[width=55mm]{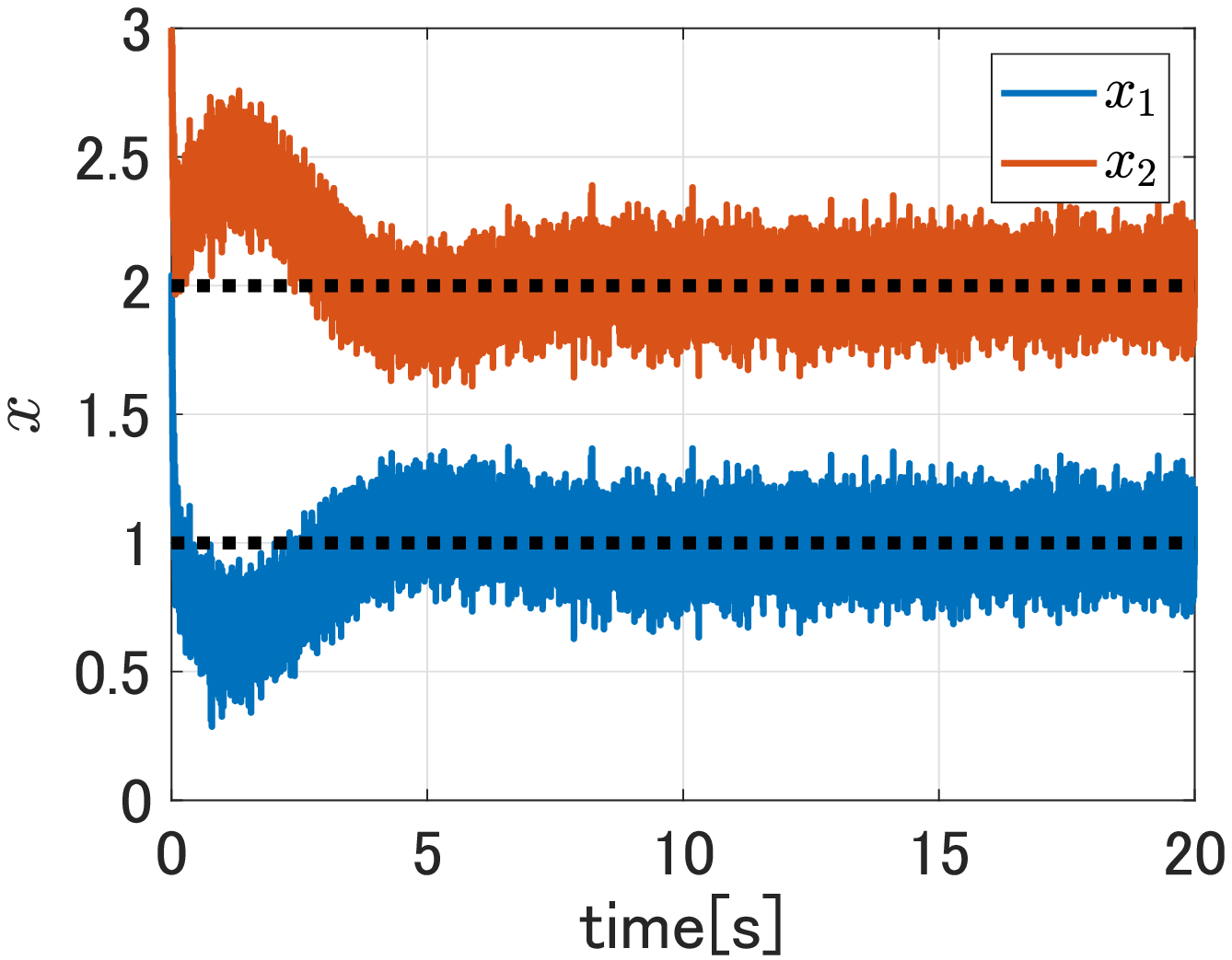}
(a)
    \end{center}
  \end{minipage}
  \begin{minipage}{.30\hsize}
\begin{center}
\includegraphics[width=55mm]{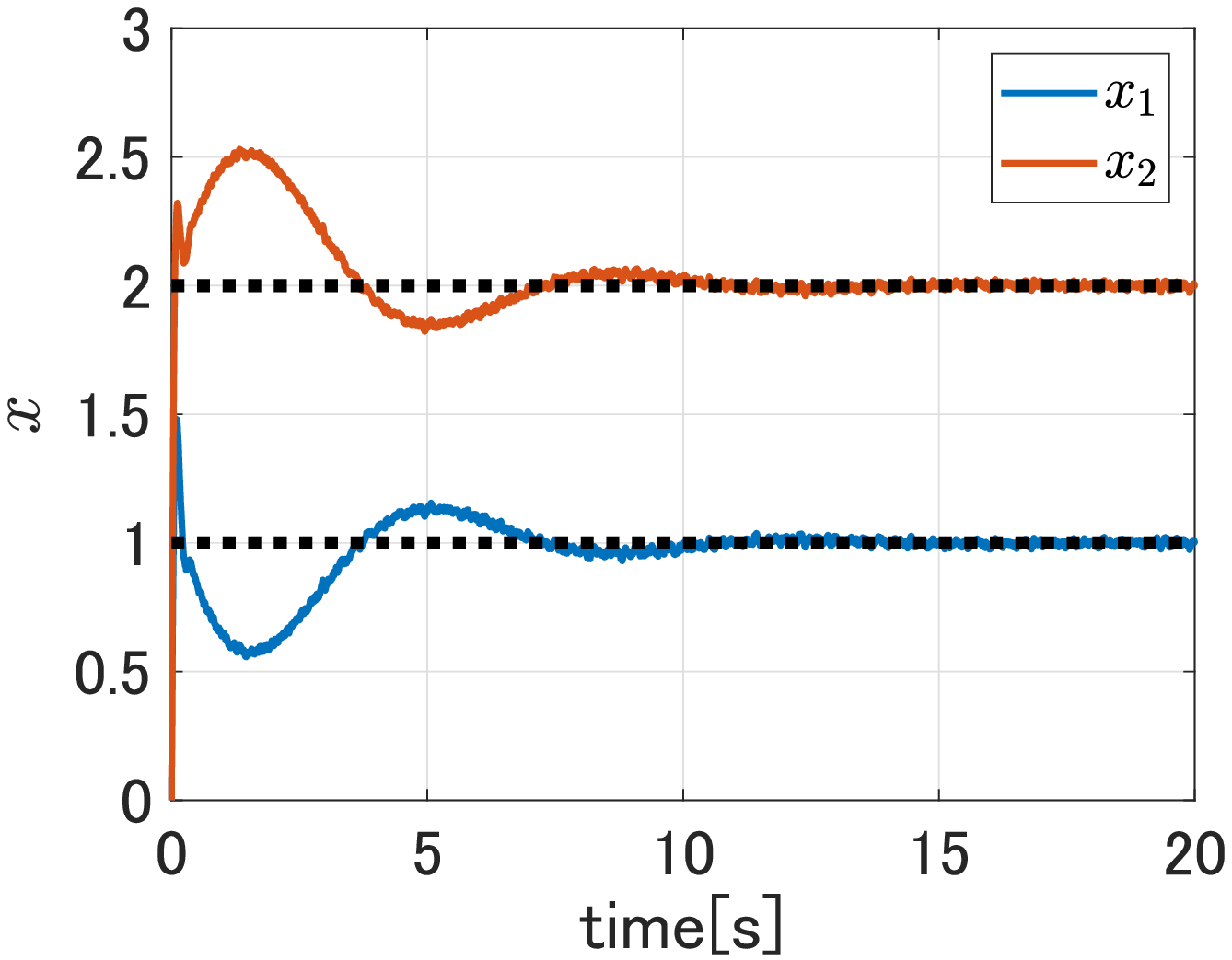}
(b)
\end{center}
  \end{minipage}
  \begin{minipage}{.30\hsize}
\begin{center}
\includegraphics[width=55mm]{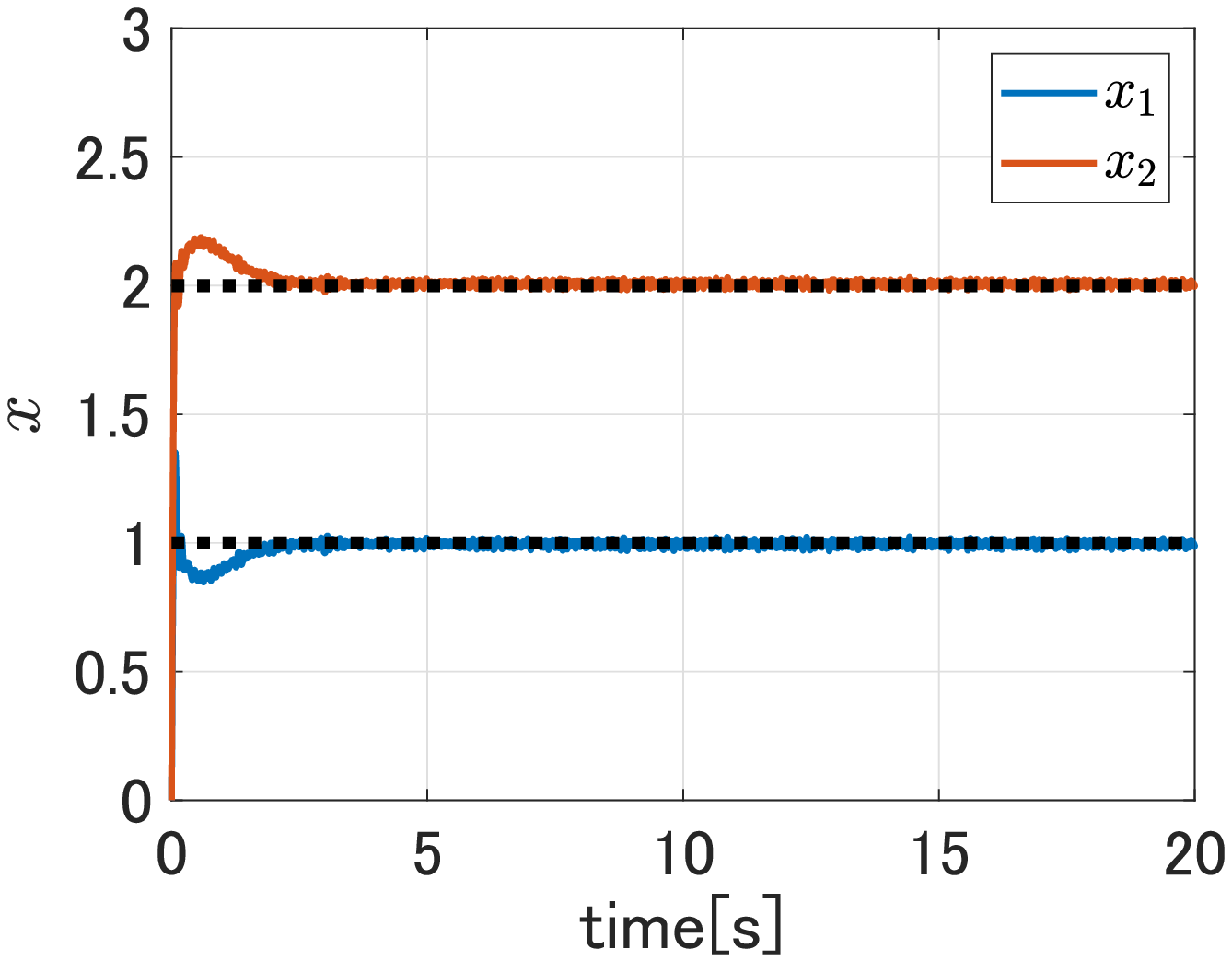}
(c)
\end{center}
  \end{minipage}
\end{tabular}
\caption{Trajectories of $x_1$ and $x_2$ with effects of noise. 
The results (a)--(c) are respectively corresponding to case 1-- case 3. 
}
\label{fig:case_n}
\end{center}
\end{figure*}

We run the algorithms with the above transfer functions
while adding the zero mean Gaussian noises with frequency components greater than 10rad/s to $\theta$.
The initial states are set as $\xi(0) = 0, \rho(0) = 0$ for all cases. 
We see from \refig{case_n}(a) that the algorithm of case 1, namely \cite{RC2015}, is heavily affected by the noise. 
On the other hand, we also see that the effects are drastically reduced in \refig{case_n}(b) and (c), where the corresponding dynamics 
have strictly proper $M_i(s)$ and $G^+_l(s)$.
Comparing (b) and (c), it is also confirmed that the convergence speed can be improved by appropriately supplying zeros to these transfer functions.

\section{Conclusion}
\label{sec:conc}

In this paper, we have presented
a generalized primal-dual dynamics based on the concept of passivity.
We have then proved asymptotic optimality for general convex optimization
with a not necessarily strict convex cost function by
supplying at least one stable zero to one of the integrators
in the dynamics based on the passivity paradigm together with the invariance principle for Carath\'eodory systems.
The present algorithm has also been shown to generalize existing augmented Lagrangian-based primal-dual dynamics.
We have then demonstrated the benefit of the present generalization.


\bibliographystyle{plain}        
\bibliography{autosam}


\appendix
\section{Proof of lemmas} 

\subsection{Proof of \relem{pasv_primal}}
\label{app:p_primal}

The Lie derivative of $S_i$ along \req{alg_tfcnx},
denoted by $\cL_P S_i$, is given as
\begin{align*}
\cL_P S_i &= (\xi_{i1} - x_i^*)v_i +
\sum_{k=2}^{n_i} {\xi}_{ik} v_i
- \sum_{k=2}^{n_i} \delta_{ik} \xi_{ik}^2
\nonumber\\
&=\left(\sum_{k=1}^{n_i}\xi_{ik} - x_i^*\right)v_i - \sum_{k=2}^{n_i} \delta_{ik} \xi_{ik}^2
\nonumber\\
&= \widetilde{x}_i v_i - d_i v_i^2  - \sum_{k=2}^{n_i} \delta_{ik} \xi_{ik}^2,
\end{align*}
where $\delta_{ik} := a_{ik}/c_{ik} > 0$.
Accordingly, it follows that
\begin{align}
\cL_P S = \widetilde{x}^\top v - \sum_{i=1}^{n} \left ( d_i v_i^2 + \sum_{k=2}^{n_i} \delta_{ik} \xi_{ik}^2 \right ) .
\label{eq:ineq_l2_1}
\end{align}
Due to \req{alg_signv} and the definition of $\widetilde{u}$,
we have $v = -(\nabla f(x) - \nabla f(x^*) ) + \widetilde{u}$ and hence
\begin{align}
\widetilde{x}^\top v = - \widetilde{x}^\top (\nabla f(x) - \nabla f(x^*) ) + \widetilde{x}^\top \widetilde{u} \leq  \widetilde{x}^\top \widetilde{u} 
\label{eq:ineq_l2_2}
\end{align}
holds because of \req{ipassive}. 
From \req{ineq_l2_1} and  \req{ineq_l2_2} , we have
\begin{align}
\cL_P S \leq \widetilde{x}^\top \widetilde{u} - \sum_{i=1}^{n} \left ( d_i v_i^2 + \sum_{k=2}^{n_i} \delta_{ik} \xi_{ik}^2 \right ) \leq  \widetilde{x}^\top \widetilde{u} .
\label{eq:lem2}
\end{align}
This completes the proof of \relem{pasv_primal}.

\subsection{Proof of \relem{pasv_duale}}
\label{app:p_duale}

We take the notation $\widetilde{\mu} := \mu - \mu^*$. 
In the same way as \relem{pasv_primal},
the Lie derivative of $W$ along \req{alg_tfcnm}, denoted by $\cL_E W$,
satisfies
\begin{align*}
\cL_E W = h^\top \widetilde{\mu} - \sum_{j=1}^{r}\left ( \bar{d}_j h_j^2 + \sum_{q=2}^{r_i} \bar{\delta}_{jq} \zeta_{jq}^2 \right ),
\end{align*}
where $\bar{\delta}_{jq} := \bar{a}_{jq}/\bar{c}_{jq} > 0$.
Since $A x^* = b$, we have $h = A x - b = A \widetilde{x}$, and hence $h^\top \widetilde{\mu} = \widetilde{x}^\top A^\top \widetilde{\mu} = \widetilde{x}^\top \widetilde{\psi}$ holds. 
We thus obtain 
\begin{align}
\cL_E W &=  \widetilde{x}^\top \widetilde{\psi} - \sum_{j=1}^{r}\left ( \bar{d}_j h_j^2 + \sum_{q=2}^{r_i} \bar{\delta}_{jq} \zeta_{jq}^2 \right ) 
\leq 
\widetilde{x}^\top \widetilde{\psi} .
\label{eq:lem4}
\end{align}
This completes the proof of \relem{pasv_duale}.

\subsection{Proof of \relem{pasv_duali}}
\label{app:p_duali}

We take the notation $\widetilde{\lambda} := \lambda - \lambda^*$. 
The Lie derivative of $U_l$ along \req{alg_tfcnl}, denoted by $\cL_I U_l$, is given by 
\begin{align}
\cL_I U_l = -\frac{1}{\hat{c}_{l1}}\lambda_l^*[\hat{c}_{l1}w_l]_{\rho_{l1}}^+ +
\sum_{p=1}^{m_l} \frac{1}{\hat{c}_{lp}} {\rho}_{lp} [- \hat{a}_{lp} \rho_{lp} + \hat{c}_{lp}w_l]^+_{\rho_{lp}},
\label{eq:hatanaka_edit1}
\end{align}
where $\hat{a}_{l1} = 0$. 
Now, the equation
\begin{align}
{\rho}_{lp} [- \hat{a}_{lp} \rho_{lp} + \hat{c}_{lp}w_l]^+_{\rho_{lp}} 
= {\rho}_{lp} (- \hat{a}_{lp} \rho_{lp} + \hat{c}_{lp}w_l).
\label{eq:hatanaka_edit2}
\end{align}
holds for any $p=1,\dots,m_l$ since
${\rho}_{lp} = 0$ must hold in the case of
$[- \hat{a}_{lp} \rho_{lp} + \hat{c}_{lp}w_l]^+_{\rho_{lp}} \neq - \hat{a}_{lp} \rho_{lp} + \hat{c}_{lp}w_l$ from the definition of $[\cdot]^+_{*}$.
Suppose now that $[\hat{c}_{l1}w_l]_{\rho_{l1}}^+ \neq \hat{c}_{l1}w_l$.
Then, $[\hat{c}_{l1}w_l]_{\rho_{l1}}^+ = 0$ 
and $\hat{c}_{l1}w_l < 0$ must hold and hence
\begin{align}
0 = -\lambda_l^*[\hat{c}_{l1}w_l]_{\rho_{l1}}^+ \leq 
-\lambda_l^*\hat{c}_{l1}w_l
\label{eq:hatanaka_edit3}
\end{align}
because of $\lambda_l^*\geq 0$.
The inequality in (\ref{eq:hatanaka_edit3}) holds
in the case of $[\hat{c}_{l1}w_l]_{\rho_{l1}}^+ = \hat{c}_{l1}w_l$.
Substituting (\ref{eq:hatanaka_edit2}) and (\ref{eq:hatanaka_edit3})
into (\ref{eq:hatanaka_edit1}) yields
\begin{align}
\cL_I U_l &\leq (\rho_{l1}-\lambda_l^*)w_l +
\sum_{p=2}^{m_l}{\rho}_{lp} (- \hat{\delta}_{lp} \rho_{lp} + w_l),
\label{eq:hatanaka_edit4}
\end{align}
where $\hat{\delta}_{lp} := \hat a_{lp}/\hat c_{lp}>0$.
(\ref{eq:hatanaka_edit4}) is further rewritten as
\begin{align*}
\cL_I U_l 
& \leq \left(\sum_{p=1}^{m_l}\rho_{lp} - \lambda_l^* \right)w_l
- \sum_{p=2}^{m_l}\hat{\delta}_{lp}\rho_{lp}^2
\nonumber\\
&= \widetilde{\lambda}_l w_l - \hat d_l\max\{0,w_l\}w_l
- \sum_{p=2}^{m_l}\hat{\delta}_{lp}\rho_{lp}^2
\nonumber\\
&= \widetilde{\lambda}_l w_l - \hat d_l\max\{0,w_l\}^2
- \sum_{p=2}^{m_l}\hat{\delta}_{lp}\rho_{lp}^2,
\end{align*}
where $\widetilde{\lambda}_l$ is the $l$-th element of
$\widetilde{\lambda}$.
Accordingly, $\cL_I U$ satisfies 
\begin{align*}
\cL_I U &\leq \widetilde{\lambda}^\top g(x) - \sum_{l=1}^{m}  \left( \hat{d}_l \max \{ 0, w_l \}^2  + \sum_{p=2}^{m_l} \hat{\delta}_{lp}\rho_{lp}^2  \right) .
\end{align*}
Here, we focus on $\widetilde{\lambda}^\top g(x)$. 
Since $\lambda \geq 0$, $g(x^*) \leq 0$ and $\lambda^* \circ g(x^*) = 0$, we obtain 
\begin{align}
\widetilde{\lambda}^\top g(x) &= \widetilde{\lambda}^\top ( g(x) - g(x^*) ) + \lambda^\top g(x^*) - (\lambda^*)^\top g(x^*)
\nonumber\\
&= \widetilde{\lambda}^\top ( g(x) - g(x^*) ) + \lambda^\top g(x^*)
\nonumber\\
&\leq \widetilde{\lambda}^\top ( g(x) - g(x^*) )
= 
\sum_{l=1}^{m} \widetilde{\lambda}_l  ( g_l(x) - g_l(x^*) ) .
\label{eq:lem6_tlg}
\end{align}
From convexity of $g_l$, we have 
$g_l (x) - g_l(x^*) \leq \widetilde{x}^\top \nabla g_l (x)$ and $g_l(x^*) - g_l(x) \leq - \widetilde{x}^\top \nabla g_l (x^*)$. 
Using these inequalities with $\lambda_l \geq 0$ and $\lambda^*_l \geq 0$, we have 
\begin{align}
 & \widetilde{\lambda}_l  ( g_l(x) - g_l(x^*) ) 
 \nonumber\\
 & \qquad \quad = \lambda_l  ( g_l(x) - g_l(x^*) ) + \lambda^*_l ( g_l(x^*) - g_l(x) )
 \nonumber\\
 & \qquad \quad \leq \lambda_l  \widetilde{x}^\top \nabla g_l (x) - \lambda^*_l  \widetilde{x}^\top \nabla g_l (x^*) .
 \label{eq:lem6_lnglx}
\end{align}
From \req{lem6_tlg} and \req{lem6_lnglx}, 
\begin{align*}
\widetilde{\lambda}^\top g(x) \leq \widetilde{x}^\top \left( \sum_{l=1}^{m}  \lambda_l \nabla g_l (x) - \lambda^*_l \nabla g_l (x^*) \right) = \widetilde{x} ^\top \widetilde{\eta}
\end{align*}
holds. 
From these results, we obtain 
\begin{align}
\cL_I U &\leq \widetilde{x}^\top \widetilde{\eta} - \sum_{l=1}^{m}  \left( \hat{d}_l \max \{ 0, w_l \}^2  + \sum_{p=2}^{m_l} \hat{\delta}_{lp} \rho_{lp}^2  \right)
\nonumber\\
 &\leq \widetilde{x}^\top \widetilde{\eta} . 
 \label{eq:lem6}
\end{align}
This completes the proof of \relem{pasv_duali}.


                                        
\end{document}